\documentclass[showpacs,preprint]{revtex4}

\usepackage{graphicx}
\usepackage{dcolumn}
\usepackage{bm}

\begin{document}
\title{Doppler effects in a left-handed material: a first-principle theoretical study}
\author{Sanshui Xiao}
\email{sanshui@imit.kth.se, }
\author{Min Qiu}
\affiliation{Laboratory of Optics, Photonics and Quantum
Electronics, Department of Microelectronics and Information
Technology, Royal Institute of Technology (KTH), Electrum 229,
16440 Kista, Sweden.\\
Tel: 046-8-7904079; Fax: 046-8-7904090}

\date{\today}



\begin{abstract}
The Doppler effects for the reflected wave from a moving media are
systemically analyzed in this paper. The theoretical formula for
the Doppler shift in the left-handed material, which is described
by Drude's dispersion model, is presented. This formula is
examined by first-principles numerical experiments, which are in
agreement with the theoretical results.
\end{abstract}
 \maketitle

\section{INTRODUCTION}
The Doppler effect is a well-known phenomenon that the frequency
of a wave emitting from a moving object is shifted from the source
frequency, originally discovered by Christian Doppler in 1843
\cite{Doppler1843P1,Papasbook}. Conventional Doppler shift tells
us that the measured frequency will increase when a source and
observe approach each other. In virtue of this effect, an immense
variety of novel applications are widely established including
weather and aircraft radar system, blood flow measurement, and
discovery of new planets and binary stars. The inverse Doppler
effect refers to frequency shifts that are in the opposite sense
to those mentioned above. It has been observed in a
dielectric-loaded stripline cyclotron-resonance maser oscillator
\cite{Einat1997P1}, the moving shock wave propagating through a
photonic crystals \cite{Reed2003P1} and the moving boundary in an
electrical transmission line \cite{Seddon2003P1}. In fact, back to
1968, the inverse Doppler effect has been already predicted to
occur in a left-handed material (LHM) with simultaneously negative
permittivity and permeability \cite{Veselago1968}. Pendry \emph{et
al.} discovered a microwave-plasma thin-wire structure exhibiting
negative permittivity below the electronic plasma frequency
 \cite{Pendry1996P1} and a magnet-free split-ring resonator
structure exhibiting negative permeability below the magnetic
plasma frequency \cite{Pendry1999P1}. With the suggestion of
Pendry's theory, the first experimental verification of a LHM was
demonstrated \cite{Shelby2001}, which has recently received much
attention in the literature
\cite{Smith2002P1,Smith2003P1,Xiao2004P2,Yen2004P1,Xiao2004P3}.
However, up to our knowledge, inverse Doppler effect in a
left-handed material has not been demonstrated experimentally yet,
not even by numerical experiments.

In this paper, we will study Doppler effects for the reflected
wave from a moving media and present theoretical formula for the
Doppler shift, which will be examined by first-principles
numerical experiments using the finite difference time domain
(FDTD) method.

\section{MODEL AND THEORETICAL FORMULA}
Consider a plane electromagnetic wave propagation in medium
$\textit{I}$ and incident at an angle $\theta_1$ on an idealized
moving medium $\textit{II}$. The incident wave have an angular
frequency $\omega_1$ and a propagation vector $\mathbf{k}_1$ (In a
LHM, the direction of group velocity, same as that of energy flux,
is often antiparallel to phase velocity). To calculate Doppler
shift effectively for the reflected wave in such an inertial
system, a primed coordinate frame embedded in the moving medium is
introduced. The primed frame moves with the same velocity
$\mathbf{v}$ as the moving medium relative to the unprimed
(original) coordinate system. Here we keep the axes of the primed
coordinate system parallel to corresponding axis of the unprimed
system. Therefore, we can treat the space components of both the
primed and unprimed system as vectors in a common
three-dimensional space.

If the primed frame is moving with a velocity $\mathbf{v}$ with
respect to the unprimed, one can relate $\mathbf{r'}$ and $t'$
with $\mathbf{r}$ and $t$ through a pure Lorentz transformation
(no rotation of the space axes) expressed in three-dimensional
notation (see Podolsky and Kunz \cite{Podolskybook}),
\begin{eqnarray}\label{}
\mathbf{r'}=\mathbf{\Phi(v) r}-\gamma \mathbf{v} t, t'=\gamma
(t-(1/c ^2)\mathbf{v^T r}),
\end{eqnarray}
where
\begin{eqnarray}
\gamma=(1-v^2/c^2)^{-1/2},
\mathbf{\Phi(v)}=\mathbf{1}+(\gamma-1)\mathbf{\hat{v}}\mathbf{\hat{v}^T},
\end{eqnarray}
and where $\mathbf{\hat{v}}=\mathbf{v}/|\mathbf{v}|$ is the unit
vector in the direction of $\mathbf{v}$ and $\mathbf{\hat{v}^T}$
is the transpose of $\mathbf{\hat{v}}$. Consider a $4$-vector
expressed in the Minkowski notation by $(\mathbf{k},i\omega/c)$.
Based on the theorem of invariant phase of an electromagnetic wave
for the transformation, one can easily obtain
\begin{eqnarray}\label{frequ}
\mathbf{k'}=\mathbf{\Phi(v)k}-\gamma\mathbf{v}(\omega/c^2),
\omega'=\gamma (\omega-\mathbf{v^T k}),
\end{eqnarray}
which is the Doppler shift in going to the primed frame.

To obtain a solution to the problem posed above it is convenient
to firstly consider in the primed frame, in which the medium
$\textit{II}$ is at rest while the background moves with a
velocity $-\mathbf{v}$. In the primed frame there is no Doppler
shifts at the interface, we can set
\begin{eqnarray}
\omega'_1=\omega'_2 \label{eqn2}
\end{eqnarray}
as the required generalization, where $\omega'_1,\omega'_2$ is the
frequency of incident and reflected wave in the primed frame.
Using the inverse of Eq. (\ref{frequ}), one can replace Eq.
(\ref{eqn2}) and then obtain
\begin{eqnarray}
\omega_1 g_1= \omega_2 g_2,
\end{eqnarray}
where
\begin{eqnarray}
g_i=1- \mathbf{v^T k_i} /\omega_i.
\end{eqnarray}
Finally, one can obtain the reflected frequency in respects of the
incident frequency as
\begin{eqnarray}
\frac{\omega_2}{\omega_1}=\frac{1+n_1 v\cos\theta_1/c }{1-n'_1
v\cos\theta_2/c }, \label{doppler1}
\end{eqnarray}
where $\theta_1$ and $\theta_2$ denotes the incident and reflected
angle, and where $n_1$ and $n'_1$ denotes the refractive index of
the background for the incident and reflected wave, respectively.
In the case of moving media, the law of reflection and Snell's law
of refraction is no longer suitable at the interface. Moreover,
since the LHM is often a dispersive material, the refractive index
($n'_1$) of the background for the reflected wave is not the same
as that of the background for the incident wave ($n_1$) due to the
Doppler shift.

As a LHM is always a dispersive media, it can be described by the
following Debye model,
\begin{eqnarray}
\epsilon_r = \epsilon_\infty
\left(1+\frac{\omega_a^2}{\omega_b^2-\omega^2+i \omega \gamma}
\right ), \\
\mu_r = \mu_\infty \left(1+\frac{\omega_a^2}{\omega_b^2-\omega^2+i
\omega \gamma} \right ). \label{debye}
\end{eqnarray}
When $\epsilon_\infty=1$, $\omega_b=0$, and $\omega_a=\omega_p$,
it will be simplified to be the following Drude's lossless
dispersive model \cite{Cummer2003P1} for the relative permittivity
and permeability of the LHM
\begin{eqnarray}
\epsilon_r (\omega)=
 1-\frac{\omega_{p}^2}{\omega^2} , \label{drude1} \\
\mu_r(\omega)=1-\frac{\omega_{p}^2}{\omega^2},
 \label{drude2}
\end{eqnarray}
where $\omega_{p}$ is the plasma frequency or the magnetic
resonance frequency. The permittivity and permeability take
negative values for frequencies below $\omega_{p}$, which satisfy
$\epsilon_r (\omega_1) = \mu_r (\omega_1) = -1$ at frequency
$\omega_1=\omega_{p}/\sqrt{2}$. Combining Eq. (\ref{doppler1})
with the definition of negative refractive index in the LHM
($n=-\sqrt{\epsilon_r(\omega)\mu_r(\omega)}$), we can obtain the
Doppler shift theoretically for the reflected wave
\begin{eqnarray}
\omega_2=\frac{\omega_1(1+n_1v/c)+\sqrt{[\omega_1(1+n_1v/c)]^2-4\omega^2_{p1}(1-v/c)v/c}}{2(1-v/c)},
\label{lhmdoppler1}
\end{eqnarray}
with $n_1=1-\omega^2_{p1}/\omega^2_1$. Here $n_1$ and
$\omega_{p1}$ is the refractive index of the background for the
incident wave and the plasma or magnetic resonance frequency for
the background, respectively. Using the above formula, the Doppler
shift can be easily obtained by an analytical way if we know the
incident frequency and the parameters in each medium. Absorption
will not exit for this idealized model for the LHM. Moreover, in
this paper, one is interested with the frequency shift of the
reflected wave, not the intensity of the reflected waves.

\begin{figure}[h]
\includegraphics[width=3.0in]{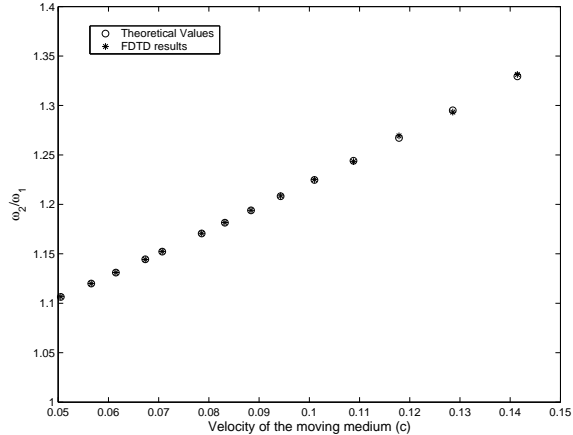}
\caption{\label{Fig1} Doppler shifts for the reflected wave from
the moving medium in air, which is a conventional material with a
refractive index of $n_2=3$. The open circle markers represent the
values predicted by the Doppler formula, and the asterisk markers
represent the numerically simulated values obtained by the FDTD
method. }
\end{figure}

\section{NUMERICAL SIMULATIONS AND RESULTS}
Next, we will give some numerical examples to examine the Doppler
formula by first-principles numerical experiments using the finite
difference time domain method \cite{TafloveFDTD}. For simplicity,
we consider the $E$-polarized electromagnetic wave incident
normally on the moving media, which moves toward the source. It is
similar for the case of the media moving away from the source. We
perform firstly a FDTD simulation to study the case of two
conventional materials. The background is air and the refractive
index of the right moving medium is $n_2=3$. A line source of
continuous wave along $y$ direction is placed at the left side of
the moving media, as well as the detector. Since we only consider
the electromagnetic wave incident normally on the interface, it is
naturally chosen the periodic condition in the $y$ direction and
the perfect matched layers (PMLs) \cite{Berenger1994P1} in the $x$
direction as numerical boundary treatments. Due to stability of
the FDTD algorithm $ (\Delta t\leq 1/ ( c \sqrt{1/\Delta
x^2+1/\Delta y^2} ) )$, the moving interface always moves less
than $\Delta x$ for each time step. In all our simulations, a
technique of parameter approximation based on the first order
Lagrange approximation is used. The theoretical Doppler formula
for the conventional material is the same as that for the case of
the LHM, except frequency-independent refractive index for
conventional materials in Eq. (\ref{doppler1}). Figure \ref{Fig1}
shows the Doppler shifts for the reflected wave from the moving
media. In Fig. \ref{Fig1}, the open circle markers represent the
values predicted by the Doppler formula, and the asterisk markers
represent the numerically simulated values obtained by FDTD
method. It can be seen from Fig. \ref{Fig1} that the results of
our simulations are in good agreement with the theoretical
results. From Fig. \ref{Fig1}, one also finds that the frequency
of reflected wave becomes larger for the media moving towards the
source, which is referred as normal Doppler shift (blueshifted).

\begin{figure}[h]
\includegraphics[width=3.0in]{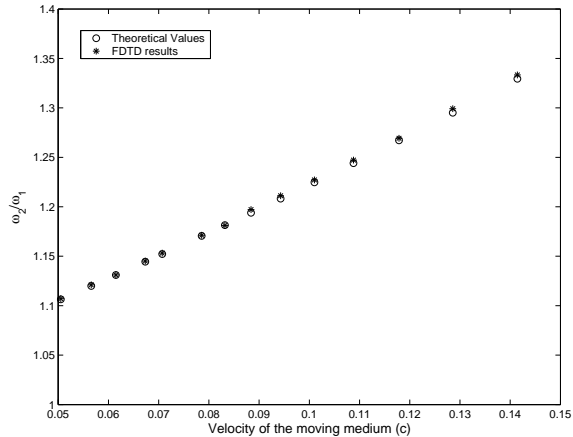}
\caption{\label{Fig2}Doppler shifts for the reflected wave from
the LHM moving in air. The refractive index of the LHM for the
incident frequency is $-1$, matched with the background (air). The
open circle and asterisk markers have the same meaning as those in
FIG. 1. }
\end{figure}
For the LHM described by Drude's model, as a special case, we
firstly consider the system with a LHM moving in air. The FDTD
method for the Drude's model is discussed in detail in
\cite{Ziolkowski2001P1}.  We choose
$\omega_1=\omega_{p2}/\sqrt{2}$, where $\omega_{p2}$ is the plasma
frequency or magnetic resonance frequency of the LHM. At that
time, the refractive index of the LHM for the incident frequency
($\omega_1$) satisfies $n_2=-1$, matched with air. It has been
shown that light can go through such an stationary air-LHM
interface without reflection for any incident angle
\cite{Pendry2000P1}. However, this is only correct for the case of
a stationary interface since the field boundary conditions at
moving interface are quite different with those at the stationary
case \cite{Podolskybook}. The Doppler shift for the reflected wave
is governed by Eq. (\ref{doppler1}) for the electromagnetic wave
incident normally on the moving interface. Results for the Doppler
shifts for the reflected wave are shown in Fig. \ref{Fig2}. The
values predicted by the Doppler formula (Eq. (\ref{lhmdoppler1}))
are presented by open circle markers, and simulation results
obtained by the FDTD method are shown by asterisk markers. From
Fig. \ref{Fig2}, one finds that simulation results are in good
agreement with the theoretical values, which are also consistent
with the results in Fig. \ref{Fig1}. It can be understood from Eq.
(\ref{doppler1}) that the Doppler shift for the reflected wave
only relates with velocity of the moving media and electromagnetic
property of the background, no any relation with the moving media.

\begin{figure}
\includegraphics[width=3.0in]{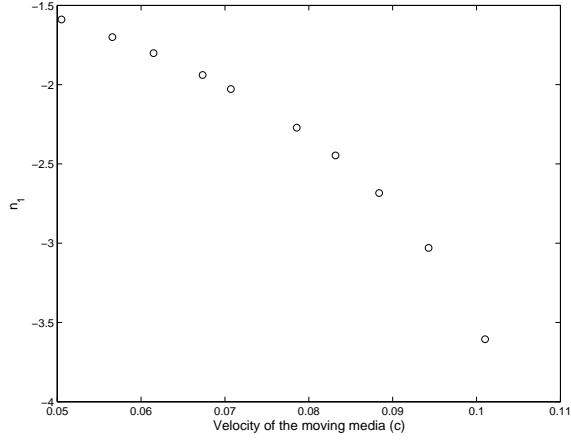}
\caption{\label{Fig3} The effective refractive index of the
background for the reflected wave as the velocity of the moving
medium increases.}
\end{figure}
\begin{figure}
\includegraphics[width=3.0in]{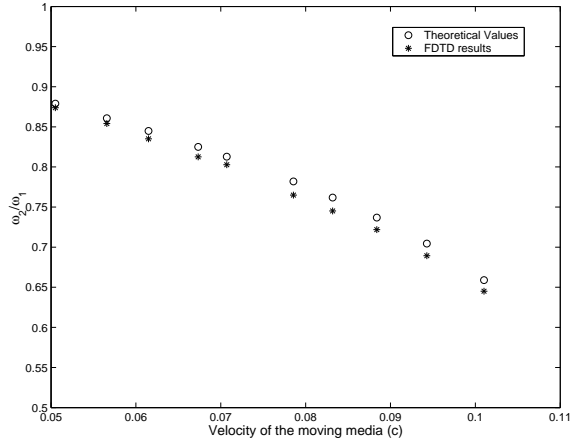}
\caption{\label{Fig4} Doppler shifts for the reflected wave from a
LHM moving in a background with another LHM. The refractive index
of the background and the moving media for the incident wave is
$-1$ and $-3$, respectively. The open circle and asterisk markers
have the same meaning as those in FIG. 1.}
\end{figure}

As stated in the introduction, the reverse Doppler effect arises
due to the negative index of refraction. The definition of Doppler
shift tells us that when a source moves inside the LHM at the
initial stationary frame, the abnormal Doppler effect will be
observed. Here we consider the system with a LHM moving in the
background with another LHM. We choose
$\omega_1=\omega_{p}/\sqrt{2}$, $\omega_{p1}=\omega_{p}$, and
$\omega_{p2}=\sqrt{2} \omega_{p}$, which means that the refractive
index of the background and the moving media for the incident
frequency is $-1$ and $-3$, respectively. Due to Doppler shift for
the reflected wave, the refractive index of the background for the
reflected wave will change. Figure \ref{Fig3} shows that the
refractive index of the background for the reflected wave varies
from $-1.59$ to $-3.60$ (Note that the refractive index of the
background for the incident frequency is $-1$). The Doppler shift
for the reflected wave from the moving media are shown in Fig.
\ref{Fig4}, in which the circle and asterisk markers represent the
theoretical values (obtained by Eq. (\ref{lhmdoppler1})) and
simulation results (obtained by the FDTD method), respectively.
Although the simulation results have a difference with theoretical
values, the numerical errors are always less than $1.5\%$. The
errors may be caused by the numerical calculation of the FDTD
method. It can be seen from Fig. \ref{Fig4} that, compared with
the results in Figs. \ref{Fig1} and \ref{Fig2} , the Doppler shift
for the reflected wave is actually an inverse Doppler effect
(redshifted). It is also in agreement with the prediction in the
LHM.

\section{CONCLUSION}
We have studied systemically the Doppler effect for the reflected
wave from a moving media and obtained theoretical formula for the
Doppler shift in the LHM, which is described by Drude's dispersive
model. We have performed first-principles numerical experiments to
examine the Doppler shifts. It has been shown that the results
obtained by our theoretical formula are in good agreement with
those obtained by numerical experiments. Inverse Doppler effect is
confirmed in the left-handed material.

This work was supported by the Swedish Foundation for Strategic
Research (SSF) on INGVAR program, the SSF Strategic Research
Center in Photonics, and the Swedish Research Council (VR) under
project 2003-5501.


\end{document}